\documentstyle[12pt]{article}
\newcommand{\beq}{\begin{equation}}
\newcommand{\eeq}{\end{equation}}

\newcommand{\beqa}{\begin{eqnarray}}
\newcommand{\eeqa}{\end{eqnarray}}

\newcommand{\ot}{\otimes}
\newcommand{\tms}{{\times}}
\newcommand{\longto}{\longrightarrow}
\newcommand{\N}{{\bf N}}
\newcommand{\Z}{{\bf Z}}

\newcommand{\R}{{\bf R}}
\newcommand{\C}{{\bf C}}
\newcommand{\F}{{\cal F}}
\newcommand{\CP}{{\C{\rm P}}}

\newcommand{\e}{{\rm e}}
\newcommand{\T}{T}
\newcommand{\cT}{c^{}_{\T}}
\newcommand{\Mgs}{\overline{\cal M}_{g,s}}
\newcommand{\Mgsd}{\overline{\cal M}_{g,s}(M,d)}
\newcommand{\Mgsod}{\overline{\cal M}_{g,s+1}(M,d)}
\newcommand{\caL}{{\cal L}}
\newcommand{\caV}{{\cal V}}
\newcommand{\parf}{f_*}
\newcommand{\tilt}{\tilde{t}}
\newcommand{\bartial}{\bar \partial}
\newcommand{\vsp}{\vspace{0.2cm}}

\begin{document}
\baselineskip=0.61cm
\normalsize

\hfill UT-694

\vspace{-0.05cm}
\hfill November, 1994

\renewcommand{\thefootnote}{\fnsymbol{footnote}}
\vspace{2.5cm}
\begin{center}
{\Large {\bf
Constraints For Topological Strings In $D\geq 1$}}

\vspace{2cm}
{\large \sc Kentaro HORI}\footnote[2]{e-mail address:
hori@danjuro.phys.s.u-tokyo.ac.jp}

\vspace{1cm}
{\large \it Department Of Physics, University Of Tokyo

\vsp
Bunkyo-ku, Tokyo 113, Japan}

\vspace{3.5cm}

{\large \bf Abstract}

\end{center}

New relations of correlation functions are found
in topological string theory;
one for each second cohomology class of the target space.
They are close cousins of
the Deligne-Dijkgraaf-Witten's puncture and dilaton equations.
When combined with the dilaton equation and the ghost number conservation,
the equation for the first chern class of the target space
gives a constraint on the topological sum
(over genera and (multi-)degrees) of partition functions.
For the $\CP^1$ model, it coincides with the dilatation constraint
which is derivable in the matrix model
recently introduced by Eguchi and Yang.

\newpage
\renewcommand{\thefootnote}{\arabic{footnote}}
\begin{center}
{\large \bf 1. Introduction}
\end{center}

\vsp
The solution of two dimensional quantum gravity \cite{BKetal} was a
striking event toward an
understanding of the nature of quantum geometry and also
of non-perturbative string theory.
Apart from the interest in its own right,
it provides a ground to test new methods and techniques
and it may become a sourse of inspiration.
Namely, it promotes further investigations.

Use of matrix integral to count the number of triagulated surfaces was
essential in the treatment.
This integral makes possible to see that the scaling fields
generate the flows of the KdV hierarchy \cite{D}.
Moreover, the Schwinger-Dyson equations lead to a system of constraints
on the partition function
--- the Virasoro constraints \cite{FKN}.
However, in gravity or string theory in
physically interesting dimensions
it is difficult to gain some definite results by
a direct generalization of the method.
It seems that we must take other approaches.

Soon after the solution of \cite{BKetal},
it is pointed out \cite{Dis,W} that
topological gravity with perturbation gives results identical to
those of the matrix model.
This was a surprise since topological gravity counts numbers
of interest in
intersection theory on the moduli space of Riemann surfaces.
Amazingly, relations of intersection numbers
derived by a purely geometric argument coincides with
the first two of the Virasoro constraints;
the puncture equation \cite{DW} gives the string equation $L_{-1}Z=0$
and the dilaton equation \cite{W1} together with dimensional consideration
gives the dilatation constraint $L_0Z=0$.
The higher constraints $L_nZ=0$ could be derived
by a rather involved treatment of quantum field theory \cite{VV}.
It is also notable that a certain matrix integral \cite{Konts}
makes transparent the relation of the intersection theory
and the KdV hierarchy.
In addition,
there exists considerable evidence \cite{Li,W3} that
the twisted $N=2$ minimal superconformal models coupled to
topological gravity (which we refer to as the minimal models)
have similar structures and are equivalent to
the physical models of matrix chains.
These successful observations invite us to take topological field theory
as an alternative method in gravity or string theory.

Topological strings are the coupled systems of topological gravity and
topological sigma models \cite{W2}.
A model is determined for each choice of compact complex or almost complex
manifold as a target space.
The $k$-th minimal model can formally be considered as
a topological string theory with the target space of
dimension $\frac{k}{k+2}$ \cite{DVV}.
It is an interesting problem to see to what extent
the structures found in the minimal models --- such as
integrable flows or as Virasoro constraints
--- can be generalized to topological string theories
in dimension $\geq 1$ (\footnote{In this paper,
by {\it dimension} of a space, we always mean its complex dimension.}).

Recently, Eguchi and Yang have proposed a matrix model for the
topological string theory with $\CP^1$ target (the $\CP^1$ model)
\cite{EY}.
The Schwinger-Dyson equations would impose
several constraints on the partition function.
A natural question to ask is whether
these can be derived by a purely geometric argument.
The string equation \cite{EY} is derivable by
the puncture equation that holds in any topological string theory \cite{DW}.
However, the dilatation constraint (associated with rescaling of matrices)
is not derivable
only from the dimensional consideration and the dilaton equation.
This is because the dimension of the moduli space
depends on the degree of the maps and the full partition function
is the sum of contributions from all genera and all degrees.
Therefore, we need other equations with degree dependence.
In this paper, we find such equations in general models.
In the $\CP^1$ model, the constraint derived from these equations
agrees with the dilatation constraint
for the matrix integral.

The rest of the paper is organized as follows.
In section 2,
we derive the new relations of correlation functions.
It turns out that they give rise to constraints on the topological sum
generalizing the first two of the Virasoro constraints.
In section 3,
we compare our constraints for the $\CP^1$ model
with the results obtained by matrix model of \cite{EY}.
Section 4 provides a field theoretical background
for the description of the correlation functions used in section 2.

\renewcommand{\theequation}{2.\arabic{equation}}\setcounter{equation}{0}
\vspace{0.7cm}
\begin{center}
{\large \bf 2. New Equations Associated With 2nd Cohomology Classes}
\end{center}

\vsp
This section contains the main result of the paper. We refer the reader
to section 4 for some field theoretical origin of the geometric
description of the amplitudes.

\vsp
The bosonic elementary fields of topological string theory consists of
a Riemannian metric on a compact oriented surface and
a map of the surface to the target space $M$.
The topological type of such fields is classified by
the genus of the surface and the degree of the map.
The degree $d$ of a map
$f:\Sigma \to M$ is defined here as the homology class
$d=f_*[\Sigma]\in H_2(M;\Z)$ of the image.
If $H_2(M;\Z)\cong \Z$, by choosing a generator $\omega\in H^2(M;\Z)$, it
is specified by an integer
\beq
\int_d \omega=\int_{\Sigma}f^*\omega.
\eeq
If $H_2(M;\Z)\cong \Z\oplus\cdots\oplus\Z$ ($r$-components), choosing a base
$\omega_1,\cdots,\omega_r$ of $H^2(M;\Z)$, it is specified by a multiple
of integers $(d_1,\cdots,d_r)$ defined by $d_i=\int_d\omega_i$.

\vsp
Physical fields in the theory are classified into
primary fields and their gravitational descendants. Primary fields are
in one to one correspondence with the de Rham cohomology classes
of the target space. We denote by $\sigma_n(\Omega)$
the $n$-th descendant of the primary field ($=\sigma_0(\Omega)$)
corresponding to $\Omega\in H^*(M;\C)$.

\vspace{0.6cm}
{\sc 2.1 Topological String Amplitudes}

\vspace{0.3cm}
The instanton calculus reduces the functional integration for
a physical amplitude
to an integration over a finite dimensional
moduli space of Riemann surfaces with holomorphic maps.
It seems that the moduli space of stable maps introduced in \cite{KM,K2}
is the natural one to appear here. We recall the definition:
\beq
\Mgsd=\Bigl\{(\Sigma,x_1,\cdots,x_s;f)\Bigr\}
\mbox{\Large /}\cong\,,
\label{defmoduli}
\eeq
A representative of an element consists of
a Riemann surface $\Sigma$ of genus $g$
(possibly with ordinary double points),
$s$-distinct marked points in $\Sigma$
and a holomorphic map $f:\Sigma \to M$
such that
every genus $0$ (resp. $\!1$) component which maps to a point
contains at least $3$ (resp. $\!1$) marked or singular points.
For each $i$, we have the evaluation map
\beq
\phi_i:\Mgsd\longto M,\quad [\Sigma,x_1,\cdots,x_s;f]\mapsto f(x_i).
\eeq
Also, the cotangent space $T^*_{x_i}\Sigma$ varies as the fibre of
a complex line bundle $\caL_{(i)}$ over $\Mgsd$.
At the point $[\Sigma,x_1,\cdots,x_s;f]$ with smooth $\Sigma$,
denoting by $\cal T$ the tangent space to $\Mgsd$,
we have the exact sequence
\beq
0\to H^0(f^*T_M)\to {\cal T}\to H^1(T_{\Sigma}\ot_{i=1}^s{\cal O}(x_i)^{-1})
\stackrel{\parf}{\longto}H^1(f^*T_M),
\label{exact}
\eeq
where $T_{\Sigma}$ and $T_M$ are the tangent bundles of
$\Sigma$ and $M$(\footnote{
In this paper, in order to simplify the notation,
we denote the sheaf of germs of holomorphic sections of
a holomorphic vector bundle $E$ by the letter $E$ itself.
}) and
the last map is induced by the homomorphism $T_{\Sigma}\to f^*T_M$;
$v^z\mapsto v^z\partial_z f^j$. Hence, the dimension of the tangent space
$\cal T$ is the sum of the dimension $\T$ of the cokernel of $\parf$ and
the following virtual dimension (the ghost number anomaly):
\beq
\int_d c_1(M)+(3-\dim M)(g-1)+s.
\label{virtdim}
\eeq

\vsp
Under these definitions, the degree $d$ contribution to
the $g$-loop $s$-point physical amplitude
is given as follows (See \S 4).
Let us take
$\Omega_i\in H^{2q_i}(M)$ ($i=1,\cdots,s$).

\vsp
(I) If $\Mgsd$ is empty or if
the ghost number $\sum(n_i+q_i)$ does not match the virtual dimension
(\ref{virtdim}), we have
\beq
\langle\sigma_{n_1}(\Omega_1)\cdots \sigma_{n_s}(\Omega_s)\rangle_{g,d}
=0.
\eeq

\vsp
(II) If the ghost number match the virtual dimension,
we have
\beq
\langle\sigma_{n_1}(\Omega_1)\cdots \sigma_{n_s}(\Omega_s)\rangle_{g,d}
=\prod_{i=1}^s n_i! \int_{\Mgsd}\cT(\caV_{g,s,d})
\prod_{i=1}^s c_1(\caL_{(i)})^{n_i}\phi_i^*\Omega_i,
\label{ampli}
\eeq
where $\caV_{g,s,d}$ is the rank $\T$ vector bundle described as follows:
At ${\cal X}=[\Sigma,x_1,\cdots, x_s;f]$
with smooth $\Sigma$, the fibre $\caV|_{\cal X}$ is the cokernel of
the map $\parf$ in (\ref{exact}).
The dual is the space of anti-ghost zero modes, i.e. the space
of holomorphic differentials
with values in $f^*T^*_M$ that are annihilated by
$df:\rho_{zi}\mapsto \rho_{zi}\partial_zf^i$.
Namely,
$\caV^*|_{\cal X}=H^0(\Sigma,\F)$ where $\F$ is the kernel of
the sheaf map $df$:
\beq
0\to \F\to K_{\Sigma}\ot f^*T^*_M\stackrel{df}{\longto}K_{\Sigma}^{\ot 2},
\label{defF}
\eeq
in which $K_{\Sigma}$ is the sheaf holomorphic differentials on $\Sigma$
(or the cotangent bundle of $\Sigma$).
This definition of $\caV^*|_{\cal X}$
is applicable to the case in which $\Sigma$ is singular,
by considering $K_{\Sigma}$ as the dualizing sheaf (\footnote{
Let $U$ be a neighborhood of a double point $x_*$ and $U_1, U_2\subset U$
be the two branches with coordinates $z, w$ ($zw=0$).
Denoting $f|_{U_a}$ by $f_a$, one can define
$df_1=z\psi\partial_zf_1^i\frac{\partial}{\partial z^i}$ and
$df_2=-w\psi\partial_wf_2^i\frac{\partial}{\partial z^i}$
where $\psi=\frac{dz}{z}=-\frac{dw}{w}$.
Then, $df$ is defined on $U$ as $df_1+df_2$ and
(\ref{defF}) makes sense.
}).

\vspace{0.4cm}
{\it Examples}

\vsp
(i) $\Mgsd$ is empty when $g=0, s\leq 2$ and $d=0$ or $(g,s,d)=(1,0,0)$.

(ii) If $\Sigma$ is smooth,
the differential $df$ determines a one (or zero) dimensional subbundle of
$K_{\Sigma}\ot f^*T_M$ with the section $df$
and hence a subbundle $\ell_f$ of $f^*T_M$.
The latter defines
the {\it normal bundle} $N_f=f^*T_M/\ell_f$ and the {\it conormal bundle}
$N_f^*=(\ell_f)^{\perp}\subset f^*T^*_M$.
Then, we have $\F=K_{\Sigma}\ot N_f^*$ and
$\caV|_{\cal X}=H^1(\Sigma,N_f)$.

(iii) The degree zero moduli space factorizes as
\beq
\Mgs(M,0)=\Mgs\times M\,,
\eeq
where $\Mgs$ is the moduli space of stable Riemann surfaces with
marked points.
Since $df=0$ for a degree zero map, $N_f$ is
the trivial bundle $\Sigma\tms T_xM$; $x=f(\Sigma)$,
and hence
\beq
\caV|_{[\Sigma,x_1,\cdots, x_s;f]}=H^0(\Sigma,K_{\Sigma})^*\ot T_xM.
\eeq
As is well known,
$H^0(\Sigma,K_{\Sigma})$ has dimension $g$ if $\Sigma$ is smooth.
If $\Sigma$ has a double point $x_*$,
the holomorphic section of $K_{\Sigma}$ is admitted to have
a simple pole at $x_*$
with opposite residues on the two branches.
Calculation using the Riemann-Roch formula gives again
$\dim H^0(\Sigma,K_{\Sigma})=g$. Anyway, $\caV_{g,s,0}$ is a vector bundle
whose rank is $g\dim M$.

\newpage
\vspace{0.4cm}
{\it Remarks}

\vsp
(i) One crucial remark is that in general the moduli space $\Mgsd$
is not smooth and it is not obvious whether (\ref{ampli}) makes sense.
However,
when $M$ satisfy a certain condition (called convexity \cite{KM}),
the smoothness (as an orbifold) for $g=0$ is proved \cite{K2} and
hence (\ref{ampli}) makes sense.
Modification will be required in some other cases
but we proceed expecting that it would not affect our main conclusion
given below
for a certain class of target spaces.

(ii) If a generic element of $\Mgsd$ has non-trivial automorphisms,
the right hand side of (\ref{ampli}) should be devided by the order of
the automorphism group. In particular, since a generic elliptic curve
with one point has order two automorphism, degree zero contribution
is given by
$\langle \sigma_n(\Omega)\rangle_{1,0}
=\frac{1}{2}n!\int c_1(\caL)^n\phi^*\Omega$.

(iii) There may be a ``pathological'' case in which (a component of)
the moduli space has dimension less than the dimension
$\int_d c_1(M) +(3-\dim M)(g-1)+T$ of each tangent space
$\cal T$. This happens for example when
$M=\CP^3$, $g=24$ and $d=14$ \cite{Mum}.
In such a case, the number of ghost zero modes is greater than
the number of moduli. Not having a good remedy at present,
we simply put zero the contribution of such component.
This automatically follows from the expression (\ref{ampli}).

\vspace{0.6cm}
{\sc 2.2 The New Equation}

\vspace{0.3cm}
We derive a new relation of correlation functions (i.e. string amplitudes)
associated with the primary field $\sigma_0(\omega)$ for each 2nd
cohomology class $\omega\in H^2(M)$. It expresses the $s+1$ point function
\beq
\langle \sigma_0(\omega)\sigma_{n_1}(\Omega_1)\cdots
\sigma_{n_s}(\Omega_s)\rangle_{g,d}
\eeq
by the sum of certain $s$ point functions.
As in the derivation of the puncture- and the dilaton equations
\cite{DW,W1},
use of the forgetful map
\beq
\pi:\Mgsod \longto \Mgsd,
\label{forget}
\eeq
is essential.
We assume that the space $\Mgsd$ is non-empty for a while and
shall deal with the empty case separately.

\vsp
We denote the image of $[\Sigma,x_0,x_1,\cdots,x_s;f]$ under $\pi$
by $[\Sigma',x_1,\cdots,x_s;f']$:
``$0$'' is the mark for the point to be forgotten. $\Sigma'$ is obtained by
contracting (if exists)
the genus $0$ and degree $0$ component of $\Sigma$ cotaining
$x_0$ and only two other marked or singular points.
We denote by $\caL_{(i)}$ (resp. $\!\caL_{(i)}'$) the line bundle
over $\Mgsod$ (resp. $\!\Mgsd$) associated with the $i$-th point
and by $\phi_i$ (resp. $\!\phi_i'$) the evaluation map of
$\Mgsod$ (resp. $\!\Mgsd$) at the $i$-th point.
As is explained in \cite{DW}, $c_1(\caL_{(i)})$ for $i\ne 0$
is different from the pull back $\pi^*c_1(\caL_{(i)}')$ but is given by
\beq
c_1(\caL_{(i)})=\pi^*c_1(\caL_{(i)}')+[D_i]\,.
\label{fundam}
\eeq
Here, $D_i$ is the divisor in $\Mgsod$ consisting of those configurations
which include a genus $0$ and degree $0$ component containing only
the $0$-th point, the $i$-th point and one node.
Also it is shown that $c_1(\caL_{(i)})[D_i]=0$.
Since $\pi$ sends the evaluation maps; $\phi_i=\phi_i'\circ \pi$, we have
\beq
\phi_i^*\Omega_i=\pi^*{\phi_i'}^*\Omega_i\,,\qquad i=1,\cdots, s\,.
\eeq
Combining these observations, we have
\beqa
c_1(\caL_{(1)})^{n_1}\phi_1^*\Omega_1\cdots
c_1(\caL_{(s)})^{n_s}\phi_s^*\Omega_s
&=&
\pi^*\!\left\{
\,c_1(\caL_{(1)}')^{n_1}{\phi_1'}^*\Omega_1\cdots
c_1(\caL_{(s)}')^{n_s}{\phi_s'}^*\Omega_s
\right\}\nonumber\\
&&+\sum_{i=1}^s\,[D_i]\,\pi^*\!\left\{
\cdots c_1(\caL_{(i)}')^{n_i-1}{\phi_i'}^*\Omega_i \cdots\right\}.
\label{pulldiff}
\eeqa
One can also show that $\caV_{g,s+1,d}$ is the pull back of
$\caV_{g,s,d}$ and hence
\beq
\cT(\caV_{g,s+1,d})=\pi^*\cT(\caV_{g,s,d})\,.
\label{pullb}
\eeq
This is essentially because the construction of $\caV_{g,\bullet,d}$
does not refer to the marked points.
However, one should be careful if $\Sigma\ne \Sigma'$.
This is the case when
(A) $[\Sigma,x_0, \cdots, x_s;f]\in[D_i]$
for some $i=1,\cdots,s$ or when
(B) $\Sigma$ includes a genus $0$ degree $0$ component containing only
$x_0$ and two nodes, say $x_*$ and $y_*$.
In each case, $H^0(\Sigma,\F)$
is canonically isomorphic to $H^0(\Sigma',\F')$
since $H^0(\CP_0^1,K)=H^0(\CP_0^1,K\ot{\cal O}(1))=0$ for (A)
and $H^0(\CP_0^1,K\ot{\cal O}(x_*)\ot{\cal O}(y_*))=\C dz/z$ for (B)
where $\CP_0^1$ is the degree $0$ component containing $x_0$ and $z$ is the
coordinate of $\CP_0^1$ with $z(x_*)=0$ and $z(y_*)=\infty$.

\vsp
The identities (\ref{pulldiff}) and (\ref{pullb}) lead to
the puncture and the dilaton equations:
\beqa
 \langle P\sigma_{n_1}(\Omega_1)\cdots
\sigma_{n_s}(\Omega_s)\rangle_{g,d}
&=&
\sum_{i=1}^s n_i
\langle \sigma_{n_i-1}(\Omega_i)\,\prod_{j\ne i}
\sigma_{n_j}(\Omega_j)\rangle_{g,d},
\label{puncteq}\\
\langle \sigma_1(P)\sigma_{n_1}(\Omega_1)\cdots
\sigma_{n_s}(\Omega_s)\rangle_{g,d}
&=&
(2g-2+s)\cdot\langle \sigma_{n_1}(\Omega_1)\cdots
\sigma_{n_s}(\Omega_s)\rangle_{g,d},
\label{dilateq}
\eeqa
where we denote $\sigma_0(1)=P$ and $\sigma_n(1)=\sigma_n(P)$.
The latter equation is due to
$\caL_{(0)}|_{{\rm fibre\, of} \pi}=K\ot_{i=1}^s{\cal O}(x_i)$ and
$c_1(\caL_{(0)})[D_i]=0$ \cite{W1}.
Finally, we perform the integration along the fibre of $\pi$ with
$\phi_0^*\omega$-insertion. Integration of the sole form $\phi_0^*\,\omega$
gives the topological number $\int_d\omega$. One can easily see
\beq
\phi_0^*\omega\,[D_i]=[D_i]\,\pi^*{\phi_i'}^*\omega.
\eeq
Hence, we have the equation
\beqa
\langle \sigma_0(\omega)\sigma_{n_1}(\Omega_1)\cdots
\sigma_{n_s}(\Omega_s)\rangle_{g,d} \label{insteq}
&=&
\int_d \omega \cdot
\langle\sigma_{n_1}(\Omega_1)\cdots\sigma_{n_s}(\Omega_s)\rangle_{g,d}\\
&&+\,\sum_{i=1}^s
n_i \langle \sigma_{n_i-1}(\omega\wedge\Omega_i)
\,\prod_{j\ne i}\sigma_{n_j}(\Omega_j)\rangle_{g,d}.
\nonumber
\eeqa
Here, $\omega\wedge\Omega_i$ is the product
in the classical cohomology ring $H^*(M;\C)$.
Note that $\sigma_0(\omega)$ counts
the degree of the maps (the first term)
as the dilaton field counts the Euler number of the surfaces, and
it also has contact interactions with other fields
(the succeeding terms) as the puncture has.

\vsp
{\it Remark}. Actually, for $(g,s,d)=(1,1,0)$ the relation (\ref{fundam})
does not hold but $[D_i]$ must be replaced by $\frac{1}{2}[D_i]$ due to the
$\Z_2$-symmetry. However, on account of the definition
of one point functions (remark (ii) in \S 2.1),
the equations (\ref{puncteq}), (\ref{dilateq})
and (\ref{insteq}) hold without modification.

\vspace{0.4cm}
{\it The Exceptional Cases}

\vsp
We consider here, as we have promised, the case in which $\Mgsd$ is empty
but $\Mgsod$ is not: a) $(g,s,d)=(0,2,0)$ and
b) $(g,s,d)=(1,0,0)$.

\vsp
a) The moduli space is just $\overline{\cal M}_{0,3}(M,0)=M$.
Hence, the gravitational descendants $\sigma_n(\Omega)$, $n>0$
decouples and we have
\beqa
\langle P\sigma_0(\Omega_1)\sigma_0(\Omega_2)\rangle_{0,0}
&=&\int_M\Omega_1\wedge \Omega_2,
\label{exc.0punc}\\
\langle \sigma_0(\omega)\sigma_0(\Omega_1)\sigma_0(\Omega_2)\rangle_{0,0}
&=&\int_M\omega\wedge\Omega_1\wedge\Omega_2.
\label{exc.0ins}
\eeqa

b) The moduli space is
$\overline{\cal M}_{1,1}(M,0)=\overline{\cal M}_{1,1}\tms M$.
Recall the example (iii) of \S 2.1.
On a smooth torus, a cotangent vector at any point determines
a holomorphic differential on the whole surface.
Even when there is a double point, this is also the case
if the chosen point is non-singular.
Thus, $H^0(\Sigma,K)$ is isomorphic to $T^*_{x_0}\Sigma$ for every
$[\Sigma,x_0]\in \overline{\cal M}_{1,1}$ and hence we have
$\caV_{1,1,0}\cong \caL_{(0)}^{-1}\ot\T_M$.
So, the Euler class is expressed as
\beq
\cT(\caV_{1,1,0})=c_{\dim M}(M)-c_1(\caL_{(0)})c_{\dim M-1}(M)+\cdots.
\eeq
Since the one point function
$\frac{1}{2}\int_{\overline{\cal M}_{1,1}}c_1(\caL_{(0)})$
of topological gravity is $\frac{1}{24}$ \cite{W1},
we have
\beqa
\langle \sigma_1(P)\rangle_{1,0}&=&\frac{1}{24}\chi(M),
\label{exc.1dil}\\
\langle \sigma_0(\omega)\rangle_{1,0}&=&
-\frac{1}{24}\int_M\omega\wedge c_{\dim M-1}(M).
\label{exc.1ins}
\eeqa
Note that $\langle P\rangle_{1,0}=0$ on dimensional ground.

\vspace{0.6cm}
{\sc 2.3 The Constraints On The Topological Sum}

\vspace{0.3cm}
Consider the theory perturbed by the physical fields. We choose a base
$\{\Omega_{\alpha}\}_{\alpha}$ of the cohomology group $H^*(M;\C)$
and denote by $t_n^{\alpha}$ the coupling constant for the field
$\sigma_n(\Omega_{\alpha})$.
The genus $g$ degree $d$ free energy is defined as
the formal series
\beq
F_{g,d}(t)=
\Bigl\langle
\exp\{\sum_{n,\alpha} t_n^{\alpha}\sigma_n(\Omega_{\alpha})\}
\Bigr\rangle_{g,d}
=\sum_{\{m_n^{\alpha}\}}\prod_{n,\alpha}
\frac{(t_n^{\alpha})^{m_n^{\alpha}}}{m_n^{\alpha}!}
\Bigl\langle\prod_{n,\alpha}\sigma_n(\Omega_{\alpha})^{m_n^{\alpha}}
\Bigr\rangle_{g,d}.
\eeq
We wish to derive constraints on the topological sum
\beq
F=F(t;\lambda,\Theta)=\sum_{g,d}\lambda^{2g-2}\e^{\int_d\Theta}F_{g,d}(t),
\eeq
where $\lambda$ is the string coupling constant and
$\Theta$ is valued in some subset of $H^2(M;\C)$.
The puncture equations (\ref{puncteq}),
the dilaton equations (\ref{dilateq}),
and the new equations (\ref{insteq}) with various insertions are
compliled into the following three respectively:
\beqa
\frac{\partial F_{g,d}}{\partial t_0^P}&=&\sum_{n,\alpha}
nt_n^{\alpha}\frac{\partial F_{g,d}}{\partial t_{n-1}^{\alpha}}
+\frac{1}{2}\sum_{\alpha,\beta}\eta_{\alpha\beta}t_0^{\alpha}t_0^{\beta}
\delta_{g,d}^{0,0},
\label{puncteq2}\\
\frac{\partial F_{g,d}}{\partial t_1^P}&=&
(2g-2)F_{g,d}+
\sum_{n,\alpha}t_n^{\alpha}\frac{\partial F_{g,d}}{\partial t_n^{\alpha}}
+\frac{\chi(M)}{24}\delta_{g,d}^{1,0},
\label{dilateq2}\\
\omega^{\alpha}\frac{\partial F_{g,d}}{\partial t_0^{\alpha}}&=&\!\!\!
\int_d\omega \,F_{g,d}+\sum_{n,\alpha,\beta}
nC_{\omega\alpha}^{\beta}t_n^{\alpha}
\frac{\partial F_{g,d}}{\partial t_{n-1}^{\beta}}
+\frac{\delta_{g,d}^{0,0}}{2}\sum_{\alpha,\beta}
C_{\omega\alpha\beta}t_0^{\alpha}t_0^{\beta}
-\frac{\delta_{g,d}^{1,0}}{24}\!\int_M\!\omega c_{\dim M-1}(M),\quad
\label{insteq2}
\eeqa
where $\eta_{\alpha\beta}=\int_M\Omega_{\alpha}\Omega_{\beta}$,
$\omega\Omega_{\alpha}=\sum \Omega_{\beta}C_{\omega\alpha}^{\beta}$
and $C_{\omega\alpha\beta}=\int_M\omega\Omega_{\alpha}\Omega_{\beta}
=\eta_{\beta\gamma}C^{\gamma}_{\omega\alpha}$.
The inhomogeneous terms have come from the exceptional contributions
(\ref{exc.0punc}), (\ref{exc.0ins}), (\ref{exc.1dil}) and (\ref{exc.1ins}).
Note also that the dimensional consideration (\S 2.1 (I))
gives the selection rule
\beq
\sum_{n,\alpha}
(n+q_{\alpha}-1)t_n^{\alpha}\frac{\partial F_{g,d}}{\partial t_n^{\alpha}}
=\Bigl( \int_d c_1(M)+(3-\dim M)(g-1)\Bigr)F_{g,d},
\label{selecteq}
\eeq
where $2q_{\alpha}$ denotes the dimension of
the base element $\Omega_{\alpha}$.

\vsp
Now we can write down two constraints on the topological sum $F$
which do not involve derivative with respect to $\lambda$ nor $\Theta$.
The puncture equation (\ref{puncteq2}) directly gives
\beq
\sum_{n,\alpha}
n \tilde{t}_n^{\alpha}\frac{\partial F}{\partial t_{n-1}^{\alpha}}
+\frac{\lambda^{-2}}{2}\sum_{\alpha,\beta}
\eta_{\alpha\beta}t_0^{\alpha}t_0^{\beta}=0.
\label{stringeq}
\eeq
The dilaton equation (\ref{dilateq2}) and
the selection rule (\ref{selecteq})
together with the new equation (\ref{insteq2}) for $\omega=c_1(M)$
give
\beqa
\lefteqn{
\sum_{n,\alpha}\bigl(n+q_{\alpha}+\mbox{$\frac{1-\dim M}{2}$}\bigr)
\tilde{t}_n^{\alpha}\frac{\partial F}{\partial t_n^{\alpha}}
+\sum_{n,\alpha,\beta}n C_{M\alpha}^{\beta}
\tilde{t}_n^{\alpha}\frac{\partial F}{\partial t_{n-1}^{\beta}}
}\nonumber\\
&+&\!\!\!
\frac{\lambda^{-2}}{2}\sum_{\alpha,\beta}
C_{M\alpha\beta}t_0^{\alpha}t_0^{\beta}
+\frac{1}{24}
\Bigl(\mbox{$\frac{3-\dim M}{2}$}\chi(M)-\int_Mc_1(M)c_{\dim M-1}(M)\Bigr)
=0.
\label{newconstr}
\eeqa
In the above expressions, we have used the parameters
$\tilde{t}_n^{\alpha}$ defined by
\beq
\tilt_n^{\alpha}=\left\{\begin{array}{ll}
t_1^P-1,& \mbox{if $(n,\alpha)=(1,P)$}\\
t_n^{\alpha},& \mbox{otherwise}.
\end{array}\right.
\label{deftilt}
\eeq
Also, we have denoted $C_{c_1(M)\alpha}^{\beta}$ etc.
by $C_{M\alpha}^{\beta}$ etc. for brevity.

\vsp
At a glance, we see that
(\ref{stringeq}) and (\ref{newconstr}) are expressed as
$L_{-1}\e^F=0$ and $L_0\e^F=0$ respectively,
using first order differenatial operators $L_0$ and $L_{-1}$
on the space of coupling constants $\{t_n^{\alpha}\}$.
We can check that the operators satisfy
\beq
[L_0,L_{-1}]=L_{-1}.
\eeq
Consider a hypothetical space $M_k$ such that $\dim M_k=\frac{k}{k+2}$,
$\chi(M_k)=k+1$, $c_1(M_k)=0$ and $H^*(M_k)$ is a $k+1$ dimensional space
generated by
$\{\Omega_{\alpha}\}_{\alpha=0,1,\cdots,k}$ with
$q_{\alpha}=\frac{\alpha}{k+1}$ and
$\eta_{\alpha\beta}=\delta_{\alpha+\beta,k}$.
Then we see that (\ref{stringeq}) and (\ref{newconstr}) are precisely the
first two of the Virasoro constraints for the $k$-th minimal model
(see \cite{Li,FKN2}).
In this sense, we may consider (\ref{stringeq}) and (\ref{newconstr}) as
the proper generalization of the string equation and
the dilatation constraint to topological strings in dimension $\geq 1$.

\renewcommand{\theequation}{3.\arabic{equation}}\setcounter{equation}{0}
\vspace{0.7cm}
\begin{center}
{\large \bf 3. The $\CP^1$ Model}
\end{center}

\newcommand{\tr}{{\rm tr}}
\newcommand{\FM}{F^{\rm M}}
\newcommand{\tiL}{\tilde{L}}

\vsp
Eguchi and Yang have recently proposed a matrix model
for the $\CP^1$ model \cite{EY}.
In this section, we derive some Schwinger-Dyson equations
for the matrix integral and compare them with the constraints
obtained in the preceding section.(\footnote{
This section is totally based on a discussion with T. Eguchi.})

\vspace{0.6cm}
{\sc 3.1 The Eguchi-Yang Model}

\vspace{0.3cm}
Let $N\in \R$ and $\nu\in \N$ be large numbers of the same order.
We consider the integration over $\nu\tms\nu$ `hermitian' matrices
\beq
Z(\nu;N)=\int d^{\nu^2}\!\!M\,\e^{N\tr V(M)},
\label{matint}
\eeq
where
\beq
V(M)=-2M(\log M-1)+\sum_{n=1}^{\infty}2t_n^PM^n(\log M-a_n)
+\sum_{n=1}^{\infty}\frac{1}{n}t_{n-1}^Q M^n,
\label{defVM}
\eeq
with $a_n=\sum_{j=1}^n\frac{1}{j}$.
Though the logarithm is not defined for a general hermitian matrix,
we could give a definition to the integral (\ref{matint})
by taking the integration region away from the set of hermitian matrices.
Namely, we replace (\ref{matint}) by
\beq
Z(\nu;N)=\frac{1}{\nu!}\int_{{\cal C}^\nu}\prod_{i=1}^{\nu}d\lambda_i\,
\prod_{i<j}(\lambda_i-\lambda_j)^2\,\e^{N\sum_{i=1}^{\nu}V(\lambda_i)},
\eeq
in which, instead of the real line,
the contour $\cal C$ is chosen so that
the integrand and the integral make sense.
(Here we have changed the normalization by multiplying a $\nu$-dependent
factor.)
However, not having a particularly good choice, we do not specify
the contour $\cal C$
and, in what follows,
we develop a formal argument by assuming that
the integration by parts does not pick up the boundary contribution.

\vsp
The orthogonal polynomials are definde by
\beqa
\psi_n(\lambda)\!&=&\!\lambda^n+\,\mbox{lower order terms},\\
\int_{\cal C}d\lambda \hspace{-0.2cm}&&\hspace{-0.7cm}\e^{N\tr V(\lambda)}
\psi_n(\lambda)\psi_m(\lambda)
=\delta_{n,m}h_n.
\eeqa
Denoting the norm squared $h_n$ by $\e^{N\phi_n}$,
the partition function is written as
\beq
Z(\nu;N)=h_0\cdots h_{\nu-1}
=h_0^{\nu}\exp\!\left\{\,\sum_{n=1}^{\nu-1}
\Bigl(\frac{\nu}{N}-\frac{n}{N}\Bigr)N^2(\phi_n-\phi_{n-1})\right\}.
\eeq
We denote by $u(\frac{n}{N};N)$ the derivative $\phi_n'$ of $\phi_n$
with repect to $\frac{n}{N}$,
in which $\phi_n$ is considered as a function of $\frac{n}{N}$ and $N$.
We also denote
$\int d\lambda\e^{NV(\lambda)}\lambda(\psi_n(\lambda))^2=v(\frac{n}{N};N)$.
The basic ansatz of \cite{EY} is that,
under the identification $\frac{n}{N}=t_0^P$,
$u$ and $v$ are the two point functions of the $\CP^1$ model
at some value of $\Theta$
\beq
\begin{array}{rcl}
u\!&=&\!\langle PP\rangle,\\[0.2cm]
v\!&=&\!\langle PQ\rangle,
\end{array}
\label{ansatz}
\eeq
with $N^{-1}$ being the string coupling constant.
Here $Q$ is the primary
field corresponding to the K\"ahler form of $\CP^1$ of volume $1$.
In particular $N^2u$ and $N^2v$ are the derivatives
of the free energy $F$ that is expanded with respect to the genus
\beq
F\left(\frac{n}{N};N\right)
=\sum_{g=0}^{\infty}N^{2-2g}F_g\left(\frac{n}{N}\right).
\eeq
Starting with this ansatz, Eguchi and Yang derived for small $g$
($g=0,1,2,3,4$) the flow equations
that can be anticipated by a general argument \cite{EYY}
(for $g=0,1$ they coincides with the topological recursion relations
of \cite{W}).
Also, just as in the matrix model for $2D$ gravities,
they derived the string equation --- the equation obtained by
differentiating our string equation
(\ref{stringeq}) with respect to $t_0^P$ and $t_0^Q$.

\vspace{0.6cm}
{\sc 3.2 The Constraints}

\vspace{0.3cm}
In this subsection,
we give further confirmation of (\ref{ansatz})
by deriving from the matrix model
(certain derivatives of) the dilatation constraint (\ref{newconstr}) and
the dilaton equation (\ref{dilateq2}) for the $\CP^1$ model.
As a preparation,
we express the ansatz in terms of the free energies of the models.
We introduce a function $f(x;N)$ such that
$N^2(\phi_n-\phi_{n-1})=\frac{1}{N}f''(\frac{n}{N};N)$.
Using the Euler-Maclaurin formula, we have the following expression
for the free energy $\FM=\log Z$ of the matrix model:
\beqa
\FM(\nu;N)\!&=&\!\nu\log h_0+\frac{1}{N}\sum_{n=1}^{\nu-1}
\left(\frac{\nu}{N}-\frac{n}{N}\right)f''\Bigl(\frac{n}{N}\Bigr)\\
&=&\!\nu\log h_0+\int^{\frac{\nu}{N}}_0\Bigl(\frac{\nu}{N}-y\Bigr)
f''(y)dy-\frac{1}{2N}\frac{\nu}{N}f''(0)\nonumber\\
&&\qquad+\sum_{r\geq 1}\frac{(-1)^{r-1}B_r}{(2r)!}\frac{1}{N^{2r}}
\left.\left(\Bigl(\frac{\nu}{N}-y\Bigr)f''(y)\right)^{(2r-1)}
\right|^{\frac{\nu}{N}}_0\nonumber\\
&=&\! N x\log h_0+f(x)-f(0)-xf'(0)-\frac{1}{2N}xf''(0)\label{relFMf}\\
&&+\sum_{r\geq 1}\frac{(-1)^{r}B_r}{(2r)!}\frac{1}{N^{2r}}
\!\left\{(2r-1)\left(f^{(2r)}(x)-f^{(2r)}(0)\right)
+xf^{(2\nu+1)}(0)\right\}
\nonumber
\eeqa
where $B_r$ are the Bernoulli numbers and $x=\frac{\nu}{N}$.
On the other hand, the ansatz (\ref{ansatz}) enables us to put
\beq
f(x;N)=F(x;N)-\frac{1}{2N}F'(x;N)+\cdots
+\frac{(-1)^r}{(r+1)!N^r}F^{(r)}(x;N)+\cdots,
\label{relfF}
\eeq
where $F$ is the free energy of the $\CP^1$ model.

\vspace{0.4cm}
{\it The Virasoro Constraints}

\vsp
Consider the differential operators
\beq
\caL_n=\sum_{i,j=1}^{\nu}\frac{\partial}{\partial M_{ij}}(M^{n+1})_{ij},
\qquad n\geq 1,
\eeq
acting on functions of the matrix $M$, where $(M^{n+1})_{ij}$ stands for
the $(i,j)$-th component of the $(n+1)$-th power of $M$.
They satisfy the commutation relations
$[\caL_n,\caL_m]=(m-n)\caL$.
When $\caL_n$ for $n\geq 0$ are applied to the weight $\e^{N\tr V(M)}$,
the results can be expressed as the responses to differential operators
on the space of coupling constants
$t_{n+1}^P, t_n^Q$:
\beq
\caL_n\e^{N\tr V(M)}=L_n\e^{N\tr V(M)},\qquad n\geq 0,
\label{difdif}
\eeq
where
\beqa
L_n\!&=&\!\frac{1}{N^2}\sum_{l=1}^{n-1}
l(n-l)
\frac{\partial}{\partial t_{l-1}^Q}
\frac{\partial}{\partial t_{n-l-1}^Q}
+\sum_{m=0}^{\infty}
\Bigl(\,m\tilt_m^P\frac{\partial}{\partial t_{m+n}^P}
+(m+n+1)t_m^Q\frac{\partial}{\partial t_{m+n}^Q}\,\Bigr)\nonumber\\
&&+2\sum_{m=0}^{\infty}
(m+n)(1-m(a_{m+n}-a_m))\tilt_m^P\frac{\partial}{\partial t_{m+n-1}^Q}
\qquad\,\, n\geq 1,\\
\noalign{\vskip0.25cm}
\mbox{and}
\quad L_0\!&=&\!\sum_{m=0}^{\infty}
\Bigl(\,m\tilt_m^P\frac{\partial}{\partial t_m^P}
+(m+1)t_m^Q\frac{\partial}{\partial t_m^Q}\,\Bigr)
+2\sum_{m=1}^{\infty}m\tilt_m^P\frac{\partial}{\partial t_{m-1}^Q}
+N^2x^2,
\eeqa
in which $x=t_0^P=\frac{\nu}{N}$ and $\tilt_n^P$ is defined
as in (\ref{deftilt}).
Due to the translational invariance of the measure $d^{\nu^2}\!M$
(or $\prod d\lambda_i$), the relation (\ref{difdif}) leads to
the following constraints on the partition function:
\beq
L_n Z=0,\qquad n\geq 0.
\label{Virconstr}
\eeq
As a consequence of the commutation relations of $\caL_n$, we find
\beq
[L_n,L_m]=(n-m)L_{n+m},\qquad n,m\geq 0.
\eeq
Hence, (\ref{Virconstr}) may be called the Virasoro
constraints.

\vspace{0.4cm}
{\it The Rescaling Constraint}

\vsp
Rescaling $N$ with $\nu=Nx$ being fixed, we obtain the equation
\beq
\Bigl(\,N\frac{\partial}{\partial N}-x\frac{\partial}{\partial x}\,\Bigr)Z
=\sum_{n=0}^{\infty}
\Bigl(\,\tilt_{n+1}^P\frac{\partial}{\partial t_{n+1}^P}
+t_n^Q\frac{\partial}{\partial t_n^Q}\,\Bigr)Z.
\label{resconstr}
\eeq
The right hand side comes out
since the action $N\tr V(M)$ depends linearly on the coupling constants
$\tilt_{n+1}^P$ and $t_n^Q$ ($n\geq 0$).
We call this the rescaling constraint.
Note that
it takes the same form as
the dilaton equation (\ref{dilateq2})
except for the inhomogeneous term $\frac{\chi}{24}$.

\vspace{0.4cm}
{\it Comparison With The Constraints For The $\CP^1$ Model}

\vsp
We are now in a position to compare
the constraints for the matrix model obtained above
with the constraints for the $\CP^1$ model.
In particular, we compare the dilatation constraint $L_0Z=0$
(the first of the Virasoro constraints (\ref{Virconstr}))
with our new constraint (\ref{newconstr}) applied to $M=\CP^1$
\beq
\Bigl(\,\tiL_0+N^2(t_0^P)^2\,\Bigr)\e^F=0,
\label{newCP1}
\eeq
where $\tiL_0=\sum_{m=0}^{\infty}
(m\tilt_m^P\frac{\partial}{\partial t_m^P}
+(m+1)t_m^Q\frac{\partial}{\partial t_m^Q})
+2\sum_{m=1}^{\infty}m\tilt_m^P\frac{\partial}{\partial t_{m-1}^Q}$.
We also compare the rescaling constraint (\ref{resconstr}) with the
dilaton equation (\ref{dilateq2})
\beq
\Bigl(\,D-N\frac{\partial}{\partial N}+\frac{1}{12}\,\Bigr)\e^F=0,
\label{dilatCP1}
\eeq
where $D=\sum_{n=0}^{\infty}(\tilt_n^P\frac{\partial}{\partial t_n^P}
+t_n^Q\frac{\partial}{\partial t_n^Q})$.
Though they have the same (similar) forms,
the comparison is a non-trivial task since $F^M$ and $F$ do not coincide
but are related by (\ref{relFMf})
with (\ref{relfF}) under the ansatz (\ref{ansatz}).
For instance, we will need the following equations:
\beqa
\left(\,\tiL_0+1\,\right)h_0\!\!&=&\!\!0,\label{dilatation1}\\
\Bigl(D-N\frac{\partial}{\partial N}\Bigl)h_0\!\!&=&\!\!0,
\label{rescale1}
\eeqa
which are the constraints $L_0Z=0$ and (\ref{resconstr})
for the special case $\nu=1$.

\vsp
We first show that the constraints (\ref{newCP1}) and (\ref{dilatCP1})
for the $\CP^1$ model lead through (\ref{relFMf}) with (\ref{relfF}) to
the dilatation and the rescaling constraints on $Z=\e^{\FM}$.
Using (\ref{newCP1}) and its derivatives with respect to $x=t_0^P$,
we find
\beq
\tiL_0f(x)=-N^2x^2+Nx-\frac{1}{3}.
\eeq
This together with its derivatives shows that
\beq
\tiL_0\FM=Nx\tiL_0\log h_0-N^2x^2+Nx.
\eeq
If we use (\ref{dilatation1}), we obtain the dilatation constraint
$L_0\e^{\FM}=0$.
On the other hand, from (\ref{dilatCP1}) we find
\beq
\Bigl(D-N\frac{\partial}{\partial N}\Bigl)\frac{1}{N^m}F^{(m)}=0\qquad
m\geq 1,
\eeq
which leads to $(D-N\frac{\partial}{\partial N})f(x)=-\frac{1}{12}$.
It then follows that
\beq
\Bigl(D-N\frac{\partial}{\partial N}\Bigl)\frac{1}{N^m}f^{(m)}=0\quad
\mbox{and}\quad
\Bigl(D-N\frac{\partial}{\partial N}\Bigl)\frac{x}{N^{m-1}}f^{(m)}(0)=0
\quad m\geq 1.
\eeq
Now, using (\ref{rescale1}),
we obtain the rescaling constraint (\ref{resconstr}).

\vsp
Conversely, if we start with $\tiL_0\FM=-\nu^2$ and
$(D-N\frac{\partial}{\partial N})\FM=0$,
by differetiating them by $x=\frac{\nu}{N}$,
we obtain recursively
(namely, order by order with respect to $N^{-1}$)
the second derivatives of our equations $\tiL_0F=-N^2x^2$ and
$(D-N\frac{\partial}{\partial N})F=-\frac{1}{12}$.

\vsp
We have thus observed that some of the matrix model results coincides with
the ones in the $\CP^1$ model obtained by geometric arguments.
It seems an interesting problem to find
further constraints for the $\CP^1$ model which correspond to
the higher Virasoro constraints $L_nZ=0$, $n\geq 1$.
In particular, we wish to know
whether there are Virasoro constraints $L_n\e^F=0$,
$n\geq -1$ for the $\CP^1$ model.
(We have already found $L_{-1}$ and $L_0$ for general models.)
A more chalenging problem is to find matrix models for
other models of topological strings so as to derive
the constraints obtained in \S 2.
For this purpose, we give in Appendix
the explicit form of the constraint for the case in which
the target space is the projective space or the grassmannian.

\renewcommand{\theequation}{4.\arabic{equation}}\setcounter{equation}{0}
\vspace{0.7cm}
\begin{center}
{\large \bf 4. Field Theoretical Background}
\end{center}

\newcommand{\Ker}{{\rm Ker}}
\newcommand{\Img}{{\rm Im}}
\newcommand{\Hol}{{\cal H}{\scriptstyle{\cal O}}\ell}
\newcommand{\bi}{\bar i}
\newcommand{\bj}{\bar j}
\newcommand{\bk}{\bar k}
\newcommand{\bl}{\bar l}
\newcommand{\bv}{\bar v}
\newcommand{\bz}{\bar z}
\newcommand{\bomega}{\bar \omega}
\newcommand{\bvarphi}{\bar \varphi}
\newcommand{\brho}{\bar \rho}
\newcommand{\bchi}{\bar \chi}
\newcommand{\J}{J^z_{\,\,\bz}}
\newcommand{\bJ}{J^{\bz}_{\,\,z}}
\newcommand{\sh}{\sharp}

\vsp
This section provides
a field theoretical background
for the description of string amplitudes given in \S 2,
especially for the emergence of the top chern class
of the vector bundle $\caV_{g,s,d}$.
Though the argument is rather standard \cite{AJ,W3,AM},
we present it in some detail because it exhibits that
the general principle also works in a model in which
gravity and matter are coupled in non-trivial way
(\footnote{
Ref. \cite{AM} takes account only of the matter degrees of freedom.
Ref. \cite{W3} deals with a certain coupled system
but the matter degrees of freedom is `discrete' in the sense that
the total moduli space is only a finite (branched) cover
of the ordinary one.}).

\vspace{0.3cm}
We start with describing briefly the elementary fields,
the classical action and the symmetries.
Bosonic elementary variables consist of a metric $g$ on the world sheet
$\Sigma$ and a map $\phi$ of $\Sigma$ to the K\"ahler manifold $M$.
The BRST transformation introduces their superpartners --- the ghosts
(\footnote{
We use $\mu, \nu, \lambda, \sigma, \cdots$ to denote the world sheet indices
and $I,J,K,L,\cdots$ for the target indices.
With respect to a chosen complex structure, we shall often use
a local complex coordinate $z$ ($\bz$) itself
for the (anti-)holomorphic index
on the world sheet and $i,j,k,l,\cdots$ ($\bi, \bj, \bk, \bl,\cdots$)
for the (anti-)holomorphic indices on the target space.}):
\beq
\begin{array}{l}
\delta g_{\mu\nu}=-2\chi_{\mu\nu}+\cdots,\quad\delta\chi_{\mu\nu}=\cdots,
\\[0.2cm]
\delta \phi^I=\chi^I,\qquad\delta\chi^I=0.
\end{array}
\eeq
In the above expression, $\cdots$ denotes the contribution of
diffeomorphism ghost and the secondary ghost
in the topological gravity multiplet (see e.g. \cite{LPW})
which will not be used in our discussion.
Also, we will eventually fix the local scale and restrict our attention to
the complex structure $J^{\mu}_{\,\,\nu}$
induced by the metric $g_{\mu\nu}$. Hence,
we shall neglect the trace part $g^{\mu\nu}\chi_{\mu\nu}$ from the start.
The classical action of the system is the sum
\beq
I=I_g+I_m,
\eeq
of the gravity part $I_g$ (e.g. eqn (5.9) in \cite{LPW}) and
the sigma model part
\beqa
\lefteqn{I_m=\frac{1}{2\pi}\int_{\Sigma}d^2z\,\Bigl\{\,
g_{i\bj}\partial_z\phi^{\bj}\partial_{\bz}\phi^i+
i\rho_{zi}(D_{\bz}\chi^i+\chi^z_{\bz}\partial_z\phi^i)+
i\rho_{\bz\bi}(D_z\chi^{\bi}+\chi^{\bz}_z\partial_{\bz}\phi^{\bi})\Bigr.}
\nonumber\\
&&\hspace{2.7cm}\Bigl.
+\chi^k\chi^{\bl}{R_{k\bl}}^{\,i}_{\,\,j}\rho^j_{\bz}\rho_{zi}
-\chi^z_{\bz}\chi^{\bz}_z\rho^i_{\bz}\rho_{zi}\,\Bigr\},\hspace{4.5cm}
\label{sgmaction}
\eeqa
where $d^2z=idzd\bz$ and
$D_{\mu}\chi^I=
(\partial_{\mu}\delta^I_J+\partial_{\mu}f^K\Gamma_{KJ}^I)\chi^J$
in which $D_Kv^I=(\partial_K\delta^I_J+\Gamma_{KJ}^I)v^J$ is
the covariant derivative with respect to the K\"ahler metric $g_{IJ}$.
The metric is used also to raise and lower the indices.
We follow the convention $[D_K,D_L]v^I={R_{KL}}^{\,I}_{\,\,J}v^J$
for the definition of the curvature tensor.
$\rho_{zi}$ (resp. $\rho_{\bz\bi}$) is a fermionic field
--- the anti-ghost ---
with values in $K_{\Sigma}\ot f^*T^*_M$
(resp. $\overline{K}_{\Sigma}\ot f^*\overline{T}^*_M$).

\vsp
This action is invariant under the BRST transformation
\beq
\begin{array}{l}
\delta \J=-2i\chi^{z}_{\bz},\quad\delta\chi^z_{\bz}=0,\qquad
\delta \bJ=2i\chi^{\bz}_{z},\quad\delta\chi^{\bz}_{z}=0,\\[0.18cm]
\delta\phi^i=\chi^i,\quad\delta\chi^i=0,\qquad
\delta\phi^{\bi}=\chi^{\bi},\quad\delta\chi^{\bi}=0\\[0.18cm]
\delta\rho_{z}^{\bi}=
i\partial_z\phi^{\bi}-\Gamma^{\bi}_{\bj\bk}\chi^{\bj}\rho_{z}^{\bk},\qquad
\delta\rho_{\bz}^{i}=
i\partial_{\bz}\phi^{i}-\Gamma^{i}_{jk}\chi^{j}\rho_{\bz}^{k}.
\end{array}
\label{BRST}
\eeq
The action (\ref{sgmaction}) and the transformation rule above is obtained
by neglecting the secondary ghosts etc. in the
corresponding expressions given in \cite{W2}.

\vspace{0.4cm}
{\it The Space Of Instantons}

\vsp
The most important fact in the quantum theory is that the WKB approximation
is exact. This is because the action is obtained by
eliminating some auxiliary fields from a BRST exact functional.
Thus, we are interested in instantons ---
configurations that minimize the action.

\vsp
Since the local scale degrees of freedom will eventually be fixed,
we look for instantons in the space
${\cal C}_{\Sigma}\tms {\rm Map}(\Sigma,M)$
of pairs of complex structures on $\Sigma$ and maps of $\Sigma$ to $M$.
Looking at the action (\ref{sgmaction}), we see that the space
of instantons is given by
\beq
\Hol=\Bigl\{(J,f)\,;\,f:\Sigma_J\to M\,\,\,\mbox{is holomorphic.}\,\Bigr\},
\eeq
where $\Sigma_J$ stands for the Riemann surface determined by
the complex structure $J$.
The quotient of this space by a certain group of diffeomorphisms is
an open subset of the moduli space of stable maps (\ref{defmoduli}).
With respect to local coordinates, the instanton condition is written as
\beq
\frac{1}{2}dx^{\mu}(\delta_{\mu}^{\,\nu}+iJ_{\,\,\mu}^{\nu})
\partial_{\nu}f^i=0.
\eeq
Taking the first order variation, we find that the tangent space to
$\Hol$ at $(J,f)$ is spanned by such $(\delta J,\delta f)$ that
\beq
\partial_{\bz}\delta f^i+\frac{i}{2}\delta \J\partial_zf^i=0.
\label{tangent}
\eeq
We can (and we do) consider the space
${\cal C}_{\Sigma}\tms{\rm Map}(\Sigma,M)$ as a complex manifold
such that $\delta^{0,1}\!\J=0$ and $\delta^{0,1}\!f^i=0$.
Since the tangential condition (\ref{tangent}) is holomorphic,
each tangent space is a complex subspace of the tangent space of
${\cal C}_{\Sigma}\tms{\rm Map}(\Sigma,M)$.
As in \S 2, we assume that the instanton space
$\Hol\subset{\cal C}_{\Sigma}\tms{\rm Map}(\Sigma,M)$
is a smooth complex submanifold.

Note that the above description naturally leads
to the description (\ref{exact}) of the tangent space to
the moduli space of stable maps.
Note also that a ghost zero mode, i.e. $\chi^z_{\bz},\chi^i$ with
$\partial_{\bz}\chi^i+\chi^z_{\bz}\partial_zf^i=0$,
gives a $(1,0)$-tangent vector of $\Hol$.

\vspace{0.6cm}
{\sc 4.1 The Instanton Calculus}

\vspace{0.3cm}
We proceed to the Gaussian approximation around an instanton.
We first note that,
after certain partial quantization of topological gravity,
we are left with the integration over the moduli space $\Mgs$ of
Riemann surfaces
with the measure provided by the integration over the fermionic fields
$\chi^z_{\bz}$, $\chi^{\bz}_z$.
Therefore, we assume in what follows that the complex structure $J$
varies in a representative family $\{J_m\}$ parametrized
by the finite moduli $m^1,\cdots, m^{3g-3+s}$ and that
$\chi^z_{\bz}$ takes values in the tangent space of the moduli space:
$\chi^z_{\bz}=\frac{i}{2}\sum\widehat{m}_a\frac{\partial}{\partial m^a}\J$.
Also, since the position of the puncture does not affect our argument,
we put $s=0$ from the start.

\vsp
Now take an instanton $(J,f)\in \Hol$ and consider the first order variation
\beqa
f&\to&f+\delta f,\\
J&\to&J+\delta J,
\eeqa
in the direction transversal to $\Hol$.
This is generated by
a smooth section $v$ of the bundle $f^*T_M$; $\delta f^i=v^i$ and
a Beltrami differential $\omega$
representing a tangent vector
$[\omega]\in H^{0,1}(T_{\Sigma})\cong H^1(T_{\Sigma})$
of the ordinary moduli space; $\delta \J=-2i\omega^z_{\bz}$.
The transversality condition is realized by the requirement
\beqa
v\!\!&\in&\!\!(H^0(f^*T_M))^{\perp}\subset \Omega^{0,0}(f^*T_M),
\label{transv}\\
\omega\!\!&\in&\!(\Ker\,f_*)^{\perp}
\subset(\bar \partial \Omega^{0,0}(T_{\Sigma}))^{\perp}.
\label{transom}
\eeqa
The expressions here are explained as follows:
For a vector bundle $E$ over $\Sigma$, we denote by $\Omega^{p,q}(E)$
the space of smooth $(p,q)$-forms with values in $E$.
For $E=T_{\Sigma}$ or $f^*T_M$, we use the metrics $g_{z\bz}$ and $g_{i\bj}$
to define a hermitian inner product on the space $\Omega^{p,q}(E)$.
The space $H^0(f^*T_M)$ of holomorphic sections is considered as
a subspace of $\Omega^{0,0}(f^*T_M)$ and $(H^0(f^*T_M))^{\perp}$ is
its orthocomplement.
The cokernel $H^{0,1}(E)$ of $\bartial:\Omega^{0,0}(E)\to \Omega^{0,1}(E)$
for $E=T_{\Sigma}$ or $f^*T_M$
is identified with the orthocomplement of the image.
As in \S 2, we have the map
$f_*:H^{0,1}(T_{\Sigma})\to H^{0,1}(f^*T_M)$
that comes from the cochain map
$f_{\sh}:\Omega^{p,q}(T_{\Sigma})\to\Omega^{p,q}(f^*T_M)$
induced by $T_{\Sigma}\to f^*T_M$;
$v^z\mapsto v^z\partial_zf^i$.

\vsp
Before starting the calculation, we remark on the complex conjugates
$\bv$ and $\bomega$ of $v$ and $\omega$.
Note first that the metrics $g_{z\bz}$ and $g_{i\bj}$
give the identification of bundles
\beq
\overline{T}_{\Sigma}\cong T^*_{\Sigma}=K_{\Sigma},\qquad
f^*\overline{T}_M\cong f^*T^*_M.
\eeq
The transversality conditions (\ref{transv}) and (\ref{transom}) can then
be stated as
\beqa
\bv\!\!&\in&\!\!(H^0(f^*T^*_M))^{\perp}
\subset\Omega^{0,0}(f^*T^*_M),
\label{transv'}\\
\bomega\!\!&\in&\!\Img \,df\subset H^0(K_{\Sigma}^{\ot 2}),
\label{transom'}
\eeqa
where $df$ is the map
$H^0(K_{\Sigma}\ot f^*T^*_M)\to H^0(K_{\Sigma}^{\ot 2})$ induced by the map
(\ref{defF}):$\rho_{zi}\to\rho_{zi}\partial_zf^i$. The equivalence of
(\ref{transom}) and (\ref{transom'}) can be seen as follows:
Under the identification
$\overline{\Omega^{0,1}(T_{\Sigma})}\cong\Omega^{0,0}(K_{\Sigma}^{\ot 2})$,
$\overline{(\bartial\Omega^{0,0}(T_{\Sigma}))^{\perp}}$ is the null
space of $\bartial\Omega^{0,0}(T_{\Sigma})$ and hence coincides with
$H^0(K_{\Sigma}^{\ot 2})$.
It then suffices to note that $df$ is the dual map of $f_*$.

\vspace{0.4cm}
{\it The Calculation}

\vsp
We expand the action $I_m$ up to quadratic terms in
$v, \bv,\omega$ and $\bomega$.
Under the variation of $J$ and $\phi$, the spaces of fields
$\chi^i$ and $\rho_{zi}$ are varied and we use the metric connection on $M$
to transport them:
\beqa
\chi^i\mbox{$\frac{\partial}{\partial z^i}$}|_f\!\!&\to&\!\!
(\,\chi^i-v^k\Gamma_{kj}^i\chi^j\,)
\mbox{$\frac{\partial}{\partial z^i}$}|_{f+\delta f},\\[0.1cm]
\rho_{zi}dz\ot dz^i|_f\!\!&\to&\!\!
\left(\,\rho_{zi}dz+v^k\Gamma_{ki}^j\rho_{zj}dz+
\omega^z_{\bz}\rho_{zi}d\bz\,\right)\!\ot dz^i|_{f+\delta f}.
\eeqa
In the above expressions,
$\frac{\partial}{\partial z^i}|_f$ (resp. $dz^i|_f$)
stand for the local sections of $f^*T_M$ (resp. $f^*T^*_M$)
associated with local coordinates $z^i$ of $M$.
By a direct calculation, we obtain the quadratic approximation
$I_m^{(2)}=\frac{1}{2\pi}I_{\rm f}+\frac{1}{2\pi}I_{\rm b}$
where
\beq
I_{\rm f}=\int_{\Sigma}d^2z\,\Bigl\{\,
i\rho_{zi}(\partial_{\bz}\chi^i+\chi^z_{\bz}\partial_zf^i)+
i\rho_{\bz\bi}(\partial_z\chi^{\bi}+\chi^{\bz}_z\partial_{\bz}f^{\bi})
+\chi^k\chi^{\bl}{R_{k\bl}}^{\,i}_{\,\,j}\rho^j_{\bz}\rho_{zi}
-\chi^z_{\bz}\chi^{\bz}_z\rho^i_{\bz}\rho_{zi}\,\Bigr\},
\eeq
and
\vspace{-0.2cm}
\beq
\begin{array}{rl}
I_{\rm b}=\displaystyle{\int_{\Sigma}}\,d^2z\,\,\!
\Bigl\{&\!\!\!-g^{\bi j}v_{\bi}D_{\bz}D_z\bv_j
+D_{\bz}v_{\bi}\bomega^{\bz}_z\partial_{\bz}f^{\bi}
+\omega^z_{\bz}\partial_zf^iD_z\bv_i
+g_{i\bj}\omega^z_{\bz}\partial_zf^i\bomega^{\bz}_z\partial_{\bz}f^{\bj}
\Bigr.\\[0.23cm]
&\Bigl.\!
-iv^i\varphi_{\bz zi}+i\bvarphi_{z\bz \bi}\bv^{\bi}
-i\omega^z_{\bz}D_z\chi^i\rho_{zi}
+i\rho_{\bz\bi}D_{\bz}\chi^{\bi}\bomega^{\bz}_z\,\Bigr\}\,+\cdots,
\end{array}
\eeq
in which
\vspace{-0.4cm}
\beqa
\varphi_{\bz zi}\!&=&\!
\chi^k\partial_{\bz}f^{\bl}{R_{k\bl}}^{\,j}_{\,\,i}\rho_{zj}
-D_z(\chi^z_{\bz}\rho_z)_i,\label{defvaphi}\\[0.1cm]
\bvarphi_{z \bz\bi}\!&=&\!
\rho_{\bz\bj}\chi^{\bk}\partial_{z}f^{l}{R_{\bk l}}^{\,\bj}_{\,\,\bi}
-D_{\bz}(\rho_{\bz}\chi^{\bz}_z)_{\bi},\hspace{2cm}
\eeqa
and $\cdots$ are terms that do not contribute
to the functional integral.

We decompose the anti-ghost $\rho$ as
$\rho^{(0)}+\rho^{(\perp)}+\rho^{(+)}$
according to the orthogonal decomposition
\beq
\Omega^{1,0}(f^*T^*_M)=H^0(K_{\Sigma}\ot N_f^*)\oplus H^0_{\perp}\oplus
D\Omega^{0,0}(f^*T^*_M),
\eeq
where $N_f^*$ is the conormal bundle (see example (ii) in \S 2.1) and
$H^0_{\perp}$ is the orthocomplement of $H^0(K_{\Sigma}\ot N_f^*)$ in
$H^0(K_{\Sigma}\ot f^*T^*_M)$. The ghosts are also decomposed as
$\chi^z_{\bz}=\chi^z_{(0)\bz}+\chi^z_{(\perp)\bz}$ and
$\chi^i=\chi^i_{(0)}+\chi^i_{(+)}$ etc. where
\beqa
\chi_{(\perp)}\in \left(\Ker f_*\right)^{\perp},&&\!\!\!\!\!
\chi_{(+)}\in (H^0(f^*T_M))^{\perp},\\
\mbox{and}\hspace{3.4cm}
\partial_{\bz}\chi^i_{(0)}+\chi^z_{(0)\bz}\partial_zf^i\!&=&\!0.
\hspace{7cm}
\eeqa
With a shift of the variables $\chi_{(+)},\bchi_{(+)}$ and
dropping terms that are irrelevant upon functional integration,
$I_{\rm f}$ can be rewritten as follows:
\beqa
\lefteqn{I_{\rm f}=\int_{\Sigma}d^2z\,\Bigl\{\,
i\rho^{(+)}\bar D\chi_{(+)}+i\rho^{(\perp)}f_*\chi_{(\perp)}
+i\brho^{(+)}D\bchi_{(+)}+i\brho^{(\perp)}\bar f_*\bchi_{(\perp)}
\Bigr.}\nonumber\\
&&\Bigl.\hspace{1.7cm}
+\chi_{(0)}\bchi_{(0)}R\brho^{(0)}\rho^{(0)}
-\chi_{(0)}\bchi_{(0)}\brho^{(0)}\rho^{(0)}\,\Bigr\}.\hspace{3.5cm}
\eeqa
Similarly, the fermionic fields $\chi,\bchi,\rho, \brho$
in $I_{\rm b}$ can be replaced by the zero mode
components $\chi_{(0)},\bchi_{(0)},\rho^{(0)}, \brho^{(0)}$.

We introduce the Green's operator $G$ for the Laplacian $\bar DD$:
\beq
G:\Omega^{1,1}(f^*T^*_M)\stackrel{\pi_{\Img\, \bar D}}{\longto}
\Img\,\bar D\stackrel{(\bar DD|)^{-1}}{\longto}\left(\Ker D\right)^{\perp}
\hookrightarrow \Omega^{0,0}(f^*T^*_M),
\label{defG}
\eeq
where $\pi_{\Img\,\bar D}$ is the orthogonal projection and $\bar DD|$
is the restriction of $\bar DD$ to $(\Ker D)^{\perp}$.
As a consequence of the replacement
$\chi\to\chi_{(0)},\rho\to\rho^{(0)}$, the two form $\varphi$
defined in (\ref{defvaphi}) is
in the image of $\bar D$. This can be seen by showing that the pairing of
$\varphi$ and any holomorphic section of $f^*T_M$ is zero.
Now we see that $I_{\rm b}$ is rewritten as
\beqa
\lefteqn{
I_{\rm b}=\int_{\Sigma}d^2z\,\Bigl\{\,
-(v+\cdots)\bar DD(\bv+\cdots)+\bvarphi G(\varphi)
\Bigr.}\\
&&\Bigl.\hspace{1.7cm}
+f_{\sh}\omega(1-DG\bar D)\bar f_{\sh}\bomega
-if_{\sh}\omega DG(\varphi)-i\omega(D\chi)\rho
-i\bvarphi G(\bar D\bar f_{\sh}\bomega)
+i\brho(\bar D\bchi)\bomega\,\Bigr\},
\nonumber
\eeqa
where
$(\bar f_{\sh}\bomega)_{zi}
=g_{i\bj}g^{\bz z}\partial_{\bz}f^{\bj}\bomega_{zz}$.
Here and in what follows,
we drop the sign `$(0)$' for the fermion zero modes.
We note that $1-DG\bar D$ is the orthogonal projection
\beq
\pi_{\cal H}:\Omega^{1,0}(f^*T^*_M)\to H^0(K_{\Sigma}\ot f^*T^*_M),
\eeq
to the kernel of $\bar D$. It is easy to see that the map
$\pi_{\cal H}\circ\bar f_{\sh}:H^0(K_{\Sigma}^{\ot 2})\to
H^0(K_{\Sigma}\ot f^*T^*_M)$ corresponds to
$\bar f_*:\overline{H^{0,1}(T_{\Sigma})}\to\overline{H^{0,1}(f^*T_M)}$
under the identification $K_{\Sigma}\cong \overline{T}_{\Sigma}$
and $f^*T^*_M\cong f^*\overline{T}_M$.
Since $\overline{(\Ker\,f_*)^{\perp}}$ corresponds to $\Img \,df$ and
the restriction of ${}^t\!f_{\sh}$ to $H^0(K_{\Sigma}\ot f^*T^*_M)$ is
the dual $df$ of $f_*$,
we see that
${}^t\!f_{\sh}\pi_{\cal H}\bar f_{\sh}$ is identified with
${}^t\!f_*\bar f_*$
and gives, by restriction, an invertible operator on
$\Img\,df$.
Using the zero mode condition for the fermions,
one can see that $D_z\chi^i\rho_{zi}+\partial_zf^iD_zG_i(\varphi)$
is a holomorphic quadratic differential.
Moreover, it seems that it is also in the image of $df$
(see the next subsection).
Now $I_{\rm b}$ can be written as
\beqa
\lefteqn{
I_{\rm b}=\int_{\Sigma}d^2z\,\Bigl\{\,
-(v+\cdots)\bar DD(\bv+\cdots)
+(\omega+\cdots){}^t\!f_{\sh}\pi_{\cal H}\bar f_{\sh}(\bomega+\cdots)
\Bigr.}\\
&&\hspace{1.8cm}\Bigl.
+\bvarphi G(\varphi)
-(\brho(\bar D\bchi)-\bvarphi G\bar D\bar f_{\sh})
\Bigl({}^t\!f_{\sh}\pi_{\cal H}\bar f_{\sh}\Bigr)^{-1}
((D\chi)\rho+{}^t\!f_{\sh}DG(\varphi))\,\Bigr\}.
\nonumber
\eeqa

\vsp
When the transversal modes $v,\omega, \rho^{(+)}$-$\chi_{(+)},
\rho^{(\perp)}$-$\chi_{(\perp)}$ (and their conjugates) are integrated out,
the bosonic and fermionic determinants cancel with each other,
and we are left with the following effective action:
\beqa
I_{\it eff}\!&=&\!\frac{1}{2\pi}\int_{\Sigma}d^2z\,\Bigl\{\,
\chi^k\chi^{\bl}{R_{k\bl}}^{\,i}_{\,\,j}\rho^j_{\bz}\rho_{zi}
-\chi^z_{\bz}\chi^{\bz}_z\rho^i_{\bz}\rho_{zi}
+\bvarphi_{z\bz}{}^i G_i(\varphi)
\Bigr.\label{Ieff}\\
&&\hspace{2cm}\Bigl.
-(\brho(\bar D\bchi)-\bvarphi G\bar D\bar f_{\sh})^{}_{z\bz}
\Bigl({}^t\!f_{\sh}\pi_{\cal H}\bar f_{\sh}\Bigr)^{-1}
\!((D\chi)\rho+{}^t\!f_{\sh}DG(\varphi))\,\Bigr\}.
\nonumber
\eeqa
We note again that the fermions in the above expression are
the zero mode components:
$\partial_{\bz}\chi^i+\chi^z_{\bz}\partial_z f^i=0$,
$\partial_{\bz}\rho_{zi}=0$ and $\rho_{zi}\partial_zf^i=0$.

\newcommand{\tilrho}{\tilde{\rho}}
\newcommand{\delrho}{\mbox{\small $\sl \Delta$}{\rho}}
\newcommand{\sGamma}{{\sl \Gamma}}
\newcommand{\snab}{{\sl \nabla}}

\vspace{0.6cm}
{\sc 4.2 Connection And Curvature Of $\caV$}

\vspace{0.3cm}
If we integrate $\e^{-I_{\it eff}}$ over the $\rho$-zero modes,
we obtain a differential form on the space $\Hol$ of instantons.
In this subsection, we show that this is the top chern class
of the dual $\caV$ of the bundle $\caV^*$ of $\rho$-zero modes
(\footnote{
As explained in \cite{W3}, whether you take $\cT(\caV^*)$
or $\cT(\caV)$ is a matter of convention. We take the latter
so that the resulting constraint (\ref{newconstr})
recover the dilatation constraint for the minimal model.
}).
To be more precise, we introduce a hermitian connection of the bundle
$\caV^*$, calculate the curvature,
and then compare it with the expression (\ref{Ieff}).

\vspace{0.4cm}
{\it The Hermitian Vector Bundle $\caV^*$}

\vsp
As the instanton $(J,f)$ varies,
the first cohomology group $H^1(N_f)$ of the normal bundle $N_f$
(see example (ii) of \S 2.1)
varies as the fibre of a bundle $\caV$ over $\Hol$.
Its dual $\caV^*$ is the bundle of $\rho$-zero modes (\footnote{
In this subsection,
we consider $\rho_{zi}$ as a {\it bosonic} field.}):
\beq
\partial_{\bz}\rho_{zi}=0,\quad\rho_{zi}\partial_zf^i=0.
\label{rhozero}
\eeq
Since each fibre $H^0(K_{\Sigma}\ot f^*T^*_M)$ is in the space
$\Omega^{1,0}(f^*T^*_M)$ with a hermitian product
(induced by the K\"ahler metric of $M$),
we can consider $\caV^*$ as a hermitian vector bundle.

\vspace{0.4cm}
{\it The Hermitian Connection Of $\caV^*$}

\vsp
We shall introduce a connection of $\caV^*$ by finding a method of parallel
transport and then show that it is hermitian.
Consider the variation $(J,f)\to(J+\delta J,f+\delta f)$ in $\Hol$;
$\partial_{\bz}\delta f^i+\frac{i}{2}\delta\J\partial_z f^i=0$.
We wish to determine how to transport $\rho\in\caV^*|_{(J,f)}$ to
an element $\tilrho\in\caV^*|_{(J+\delta J,f+\delta f)}$.
For $\tilrho$ to be a $(1,0)$-form with respect to $J+\delta J$,
it has to take the following form:
\beq
\tilrho=\Bigl(\,dz\rho_{zi}-\frac{i}{2}d\bz \delta\J\rho_{zi}
+dz\eta_{zi}\,\Bigr)\ot dz^i|_{f+\delta f}.
\eeq
We determine $\eta_{zi}$ by requiring also that $\tilrho$ is
holomorphic with respect to $J+\delta J$ and conormal with respect to
$f+\delta f$. The requirements are written down as
\beqa
\partial_{\bz}\eta_{zi}\!&=&\!-\frac{i}{2}\partial_z(\delta\J\rho_{zi}),
\label{req1}\\
\eta_{zi}\partial_z f^i\!&=&\!-\rho_{zi}\partial_z\delta f^i.
\label{req2}
\eeqa
In order to express these in covariant forms,
we consider the parallel transport
$\tilrho\in \caV^*|_{(J+\delta J,f+\delta f)}\mapsto
\tau\tilrho\in \caV^*|_{(J+\delta J,f)}$ induced by
the metric $g_{i\bj}$ of $M$:
\beq
\tau\tilrho=\Bigl(\,dz\rho_{zi}-\frac{i}{2}d\bz \delta\J\rho_{zi}
+dz\delrho_{zi}\,\Bigr)\ot dz^i|_{f},
\eeq
where $\delrho_{zi}=\eta_{zi}-\delta f^k\Gamma_{ki}^j\rho_{zj}$.
Then, the requirements (\ref{req1}) and (\ref{req2}) are written as
\beqa
D_{\bz}\delrho_{zi}\!&=&\!\varphi_{\bz zi},\label{req1'}\\
\delrho_{zi}\partial_zf^i\!&=&\!-\rho_{zi}D_z\delta f^i,\label{req2'}
\eeqa
in which $\varphi_{\bz zi}=
\delta f^k\partial_{\bz}f^{\bl}{R_{k\bl}}^{\,j}_{\,\,i}\rho_{zj}
-\frac{i}{2}D_z(\delta\J\rho_z)_i$.

\vsp
As noted in \S 4.1, if $\rho$ is a zero mode
(i.e. satisfies (\ref{rhozero}))
and if $(\delta J,\delta f)$ is a variation in $\Hol$ (a ghost zero mode),
one can show that $\varphi$ is in the image of $D_{\bz}$.
Therefore, (\ref{req1'}) can be solved by using the Green's operator $G$
for $\bar DD$, defined in (\ref{defG}), as
\beq
\delrho_{zi}=D_zG_i(\varphi)-\xi_{zi}\quad;\,\,\partial_{\bz}\xi_{zi}=0.
\eeq
We shall determine $\xi_{zi}$ so that this satisfies the second requirement
(\ref{req2'}):
\beq
\xi_{zi}\partial_zf^i=\rho_{zi}D_z\delta f^i+D_zG_i(\varphi)\partial_zf^i.
\label{rexi}
\eeq
Using again the zero mode condition of $\rho$ and $(\delta J,\delta f)$,
one can show that
the right hand side is a holomorphic quadratic differential.
Though we do not have an explicit proof,
we expect that it is in the image of
$df:H^0(K_{\Sigma}\ot f^*T^*_M)\to H^0(K_{\Sigma}^{\ot 2})$.
This is based on the fact that
there exists a solution to our problem,
namely, there exists a connection of $\caV^*$.
Let us introduce the `Green's operator' for $df$:
\beq
\sGamma:H^0(K_{\Sigma}^{\ot 2})\stackrel{\pi_I}{\longto}
\Img\,df\stackrel{(df|)^{-1}}{\longto}
(\Ker\,df)^{\perp}\hookrightarrow H^0(K_{\Sigma}\ot f^*T^*_M),
\eeq
where the projection $\pi_I$ and the orthocomplement are determined
with respect to
the inner products on $H^0(K^2)$ and $H^0(K\ot f^*T^*_M)$
inherited from $\Omega^{1,0}(K)$ and $\Omega^{1,0}(f^*T^*_M)$.
A solution of (\ref{rexi}) is given by
\beq
\xi_{zi}=\sGamma_{zi}((D\delta^{1,0}\!f)\rho+{}^t\!f_{\sh}DG(\varphi)),
\label{solxi}
\eeq
where $\delta^{1,0}f$ is the variation of in the holomorphic direction.
Thus, we have determined a method of transportation $\rho\mapsto\tilrho$.

\vsp
We can now define the covariant derivative using this transport:
\beqa
\snab\rho\!&=&\!
\tau\Bigl(\,(\rho_i+\delta\rho_i)\ot dz^i|_{f+\delta f}-\tilrho\,\Bigr)
\nonumber\\
&=&\!\Bigl(\delta\rho_i-\delta f^k\Gamma_{ki}^j\rho_j
+\frac{i}{2}d\bz\delta\J\rho_{zi}\Bigr.\\
&&\quad\Bigl.
-dzD_zG_i(\varphi)
+dz\sGamma_{zi}((D\delta^{1,0}\!f)\rho+{}^t\!f_{\sh}DG(\varphi))\,\Bigr)
\ot dz^i|_f.\nonumber
\eeqa
An important property of this connection is that it preserves the
hermitian structure of $\caV^*$.
This follows in particular from the choice (\ref{solxi}) of the solution
of (\ref{rexi}) because the image of $\sGamma$ is orthogonal to the fibre
$\Ker\,df=H^0(K_{\Sigma}\ot N_f^*)$.
This fact implies that only the $(1,1)$-form part of the
curvature $F$ is non-vanishing and the calculation
becomes relatively easy.

\vspace{0.4cm}
{\it The Curvature}

\vsp
Now we calculate the curvature:
\beqa
F\rho_i\!&=&\!F^{(1,1)}\rho_i
=\{\snab^{0,1},\snab^{1,0}\}\rho_i\nonumber\\
&=&\!
\delta f^k\delta f^{\bl}\partial_{\bl}\Gamma_{ki}^j\rho_j
-\frac{1}{4}dz\delta\bJ\delta\J\rho_{zi}\\
&&\quad-(\delta^{0,1}\!D)G_i(\varphi)
+(\delta^{0,1}\!\sGamma_i)((D\delta^{1,0}\!f)\rho+{}^t\!f_{\sh}DG(\varphi))
+D(\cdots)_i+\sGamma_{i}(\cdots).\nonumber
\eeqa
Since we will eventually take the inner product with an element of
$H^0(K_{\Sigma}\ot N_f^*)$, we can forget the terms $D(\cdots)$ and
$\sGamma_{i}(\cdots)$.

\vsp
The $(0,1)$-variation of the covariant derivative $D=dz D_z$
is directly calulated:
\beqa
(\delta^{0,1}\!D)G_i(\varphi)\!&=&\!
\delta^{0,1}\Bigl(\,
dx^{\mu}\frac{\delta_{\mu}^{\nu}-iJ_{\,\,\mu}^{\nu}}{2}
(\partial_{\nu}\delta^j_i-\partial_{\nu}f^k\Gamma_{ki}^j)\,
\Bigr)G_j(\varphi)\nonumber\\
&=&\!-dz\partial_z f^k\delta f^{\bl}{R_{\bl k}}_{\,\,i}^{\,j}G_j(\varphi)
-\frac{i}{2}dz\delta\bJ D_{\bz}G_i(\varphi).
\label{varD}
\eeqa
Taking the inner product with $\rho'\in H^0(K_{\Sigma}\ot N_f^*)$,
we have
\beq
\Bigl(\rho',(\delta^{0,1}\!D)G(\varphi)\Bigr)
=\int_{\Sigma}d^2z\bvarphi'_{z\bz}{}^{j} G_j(\varphi),
\eeq
where $\bvarphi'_{z\bz\bi}=
\brho'_{\bz\bj}\delta f^{\bl}\partial_z f^k {R_{\bl k}}^{\,\bj}_{\,\,\,\bi}
+\frac{i}{2}D_{\bz}(\brho'_{\bz}\delta\bJ)_{\bi}$.

\vsp
We have to determine the $(0,1)$-variation of the Green's operator
$\sGamma$ for $df$.
For this, we use the variational formula elaborated in \cite{AdPW},
section {\bf 3b}:
\beq
\delta\sGamma=\sGamma\delta(df)\sGamma
+\pi_K\delta(df)^{\dag}\sGamma^{\dag}\sGamma
+\sGamma\sGamma^{\dag}\delta(df)^{\dag}\pi_{I^{\perp}},
\label{varformu1}
\eeq
where $\pi_K$ and $\pi_{I^{\perp}}$ is the orthogonal projection to the
kernel and the orthocomplement of the image of $df$.
Since $\sGamma$ is valued in $(\Ker\,df)^{\perp}$,
we can neglect the first and the third terms.
To evaluate the second term, we look at the hermitian conjugate of $df$:
\beq
(\eta,(df)^{\dag}\beta)=(df(\eta),\beta)=
\int_{\Sigma}d^2z\overline{\eta_{zi}}g^{\bi j}
\Bigl(g_{j\bl}g^{z\bz}\partial_{\bz}f^{\bl}\beta_{zz}\Bigr).
\eeq
Namely, we have
\beq
(df)^{\dag}=\pi_{\cal H}\circ \bar f_{\sh},
\eeq
where we recall that $\bar f_{\sh}$ is the map
$\Omega^{1,0}(T^*_{\Sigma})\to\Omega^{1,0}(f^*T^*_M)$;
$\beta_{zz}\mapsto g_{i\bl}g^{z\bz}\partial_{\bz}f^{\bl}\beta_{zz}$
and $\pi_{\cal H}$ is the projection to $H^0(K_{\Sigma}\ot f^*T^*_M)$.
Since $df$ is the dual of $f_*$,
we have $(df)(df)^{\dag}={}^t\!f_{\sh}\pi_{\cal H}\bar f_{\sh}$
and hence
its Green's operator $\sGamma^{\dag}\sGamma$ has the restriction
\beq
\sGamma^{\dag}\sGamma
=\Bigl({}^t\!f_{\sh}\pi_{\cal H}\bar f_{\sh}\Bigr)^{-1}
\quad\mbox{on}\quad\Img\,df.
\label{relop}
\eeq
The $(0,1)$-variation of the map $\bar f_{\sh}$ is directly calculated:
\beqa
(\delta^{0,1}\bar f_{\sh})\beta\!&=&\!
\delta^{0,1}(dx^{\mu}g_{i\bl}g^{\nu\sigma}\partial_{\sigma}f^{\bl})
\beta_{\mu\nu}dz^i|_f\hspace{2cm}\nonumber\\
&=&\!
dz g_{i\bl}g^{z\bz}D_{\bz}\delta^{0,1}\!f^{\bl}\beta_{zz}dz^i|_f.
\eeqa
As for the variation of $\pi_{\cal H}$,
since it is the orthogonal projection to the kernel of the operator
$\bar D:\Omega^{1,0}(f^*T^*_M)\to\Omega^{1,1}(f^*T^*_M)$,
we can again use the variational formula of \cite{AdPW}:
\beq
\delta^{0,1}\pi_{\cal H}=-\bar D^{-1}(\delta^{0,1}\!\bar D)\pi_{\cal H}
-\pi_{\cal H}(\delta^{0,1}\!\bar D^{\dag})(\bar D^{-1})^{\dag},
\eeq
where $\bar D^{-1}$ is the Green's operator for $\bar D$.
Since $\bar D^{-1}$ is valued in $(H^0(K_{\Sigma}\ot f^*T^*_M))^{\perp}$,
we can neglect the first term in the right hand side.
Using $\bar D^{\dag}=-D*$ where $*:\Omega^{1,1}\to\Omega^{0,0}$
is the Hodge operator, we obtain
\beq
\delta^{0,1}\pi_{\cal H}
=-\pi_{\cal H}(\delta^{0,1}\!D)**^{-1}D^{-1}+\cdots
=-\pi_{\cal H}(\delta^{0,1}\!D)G\bar D+\cdots,
\eeq
in which $\delta^{0,1}\!D$ is calculated in (\ref{varD}).
Now, the inner product of $\rho'\in H^0(K_{\Sigma}\ot N_f^*)$ and
$\delta^{0,1}(df)^{\dag}\beta$ is expressed as follows:
\beqa
\Bigl(\,\rho',\delta^{0,1}(df)^{\dag}\beta\,\Bigr)
\!&=&\!\Bigl(\,\rho',\pi_{\cal H}(\delta^{0,1}\!\bar f_{\sh})\beta
+(\delta^{0,1}\pi_{\cal H})\bar f_{\sh}\beta\,\Bigr)\nonumber\\
&=&\!\int_{\Sigma}d^2z\,\Bigl\{\,
\brho'_{\bz\bl}g^{z\bz}D_{\bz}\delta^{0,1}\!f^{\bl}\beta_{zz}
-\bvarphi'_{z\bz}{}^iG_i(\bar D\bar f_{\sh}\beta)\,\Bigr\}.
\label{varformu2}
\eeqa
Using the variational formulae (\ref{varformu1}), (\ref{varformu2})
and the equality (\ref{relop}), we have
\beqa
\lefteqn{
\Bigl(\,\rho',(\delta^{0,1}\sGamma)((D\delta^{1,0}\!f)\rho
+{}^t\!f_{\sh}DG(\varphi))\,\Bigr)}\nonumber\\
&&\hspace{1cm}=\int_{\Sigma}d^2z\,
(\brho'(\bar D\delta^{0,1}\!f)-\bvarphi'G\bar D\bar f_{\sh})^{}_{z\bz}
\Bigl({}^t\!f_{\sh}\pi_{\cal H}\bar f_{\sh}\Bigr)^{-1}
\!((D\delta^{1,0}\!f)\rho+{}^t\!f_{\sh}DG(\varphi)).
\eeqa

\vsp
Gathering all we have got,
we obtain the following expression of the curvature:
\beqa
(\rho',F\rho)\!&=&\!
\int_{\Sigma}d^2z\,\Bigl\{\,
\mbox{$\frac{1}{4}$}\brho'_{\bz}{}^i\delta\J\delta\bJ\rho_{zi}-
\brho'_{\bz}{}^i\delta f^k\delta f^{\bl}{R_{k\bl}}^{\,j}_{\,\,i}\rho_{zj}
-\bvarphi'_{z\bz}{}^iG_i(\varphi)\Bigr.\\
&&\hspace{1cm}\Bigl.
+(\brho'(\bar D\delta^{0,1}\!f)-\bvarphi'G\bar D\bar f_{\sh})^{}_{z\bz}
\Bigl({}^t\!f_{\sh}\pi_{\cal H}\bar f_{\sh}\Bigr)^{-1}
\!((D\delta^{1,0}\!f)\rho+{}^t\!f_{\sh}DG(\varphi))\,\Bigr\}.\nonumber
\eeqa
We see that this coincides with $-2\pi I_{\it eff}$ under the
substitution $\delta f^i\to\chi^i,\frac{i}{2}\delta\J\to\chi^z_{\bz},
\brho'_{\bz\bi}\to \rho_{\bz\bi}$ and $\rho_{zi}\to\rho_{zi}$.
This establishes our claim.

\renewcommand{\theequation}{5.\arabic{equation}}\setcounter{equation}{0}
\vspace{0.7cm}
\begin{center}
{\large \bf 5. Concluding Remarks}
\end{center}

\vsp
We finish this paper by commenting on the physical significance of our
new constraint (\ref{newconstr}).
In the two-dimensional gravity theory,
the dilatation constraint $L_0Z=0$ together with the dilaton equation
determines the scaling behaviour of the system. In addition, it can be used
to search for multi-critical points \cite{DVVnotes}.
To see whether this holds in general topological string theory,
following \cite{DVVnotes} we take the combination
((\ref{newconstr})$+
\frac{1-\dim M}{2}$(\ref{dilateq2}))/$\delta\,+$(\ref{dilateq2})
of the constraints on $F_g=\sum_d \e^{\int_d\Theta}F_{g,d}$
where $\delta$ is a positive number.
This leads to
\beqa
x\frac{\partial F_g}{\partial x}\!&=&\!
\sum_{n,\alpha}{}'
\Bigl(\,\frac{n+q_{\alpha}}{\delta}-1\,\Bigr)
\tilt_n^{\alpha}\frac{\partial F_g}{\partial t_n^{\alpha}}
+\sum_{n,\alpha,\beta}\frac{n}{\delta}C_{M\alpha}^{\beta}
\tilt_n^{\alpha}\frac{\partial F_g}{\partial t_{n-1}^{\beta}}\nonumber\\
&&\hspace{0.5cm}
+\Bigl(\,2-\frac{\dim M-1}{\delta}\,\Bigr)(1-g) F_g
+\cdots,
\eeqa
where $x=t_0^P$ and $\sum{}'$ is the sum over all $(n,\alpha)$ except
$(0,P)$. The term `$\cdots$' is the exceptional one present only
for $g=0$ or $1$.

\vsp
When $c_1(M)=0$, we can use this to find multi-critical points:
If we tune the coupling constant as $t_1^P=1$, $t_{n_1}^{\alpha_1}\ne 0$
and other $t_n^{\alpha}=0$,
the system is on the multi-critical point with
the following string susceptibility and scaling dimension:
\beq
\gamma_{\it string}=\frac{\dim M-1}{\delta_1},\qquad
\gamma_{n,\alpha}=\frac{n+q_{\alpha}}{\delta_1},
\eeq
where $\delta_1=n_1+q_{\alpha_1}$.
Applying this to the hypothetical space $M_k$ (see the end of \S 2),
we recover the KPZ formula \cite{BDKS} for the Virasoro minimal models
coupled to gravity.

\vsp
For a target space with $c_1(M)\ne 0$, however,
we cannot expect such multi-critical behaviour
because of the `off-diagonal' term
$\sum\frac{n}{\delta}C_{M\alpha}^{\beta}
\tilt_n^{\alpha}\frac{\partial}{\partial t_{n-1}^{\beta}}$.
(One cannot `diagonalize the equation' by whatever means
since the operation
$\sigma_n(\Omega)\mapsto \sigma_{n-1}(c_1(M)\wedge\Omega)$,
$\sigma_0(\Omega)\mapsto 0$ is nilpotent.)
Hence, {\it the constraint (\ref{newconstr}) exhibits in a precise way
how the first chern class $c_1(M)$ obstructs the scaling behaviour}.

\vsp
Although we cannot expect critical behaviour for the full system,
it seems possible to find a {\it sub}system that admits multi-criticality.
Namely, we restrict our attention to those observables that decouple
from the primary field $\sigma_0(c_1(M))$:
\beq
\sigma_0(\Omega)\, \quad\mbox{and}\quad
\sigma_n(\Omega')\,\,\mbox{with}\,\,c_1(M)\wedge\Omega'=0.
\eeq
Then, we can again find multi-critical points, possibly with
logarithmic scaling violation due to the exceptional terms
on the sphere and the torus.
Though it is not at all obvious what this restriction means physically,
this does not seem a bad thing to do since the known relation
(such as the string equation and topological recursion relations)
remain closed.


\vspace{0.7cm}
\begin{center}
{\large \bf Acknowledgements}
\end{center}

\vsp
I would like to thank T. Eguchi and I. Nakamura for stimulating
and valuable discussions. Many thanks are also due to
Y. Jinzenji, S.K. Yang and Y. Yamada.

\renewcommand{\theequation}{A.\arabic{equation}}\setcounter{equation}{0}
\newpage
\vspace{0.7cm}
\begin{center}
{\large \bf Appendix}
\end{center}

\vsp
In this appendix, we give an explicit expression of the new constraint
(\ref{newconstr}):
\beq
L_0\e^F=0,
\eeq
for the case in which the target space is the complex
projective space $\CP^n$ or the complex grassmann manifold $G(k,N)$.
Though $\CP^n=G(1,n+1)$, we treat them separately.

\vspace{0.6cm}
{\sc The $\CP^n$ Model}

\vspace{0.3cm}
The cohomology ring $H^*(\CP^n)$ of $M=\CP^n$ is generated by
the K\"ahler class $\omega\in H^2(\CP^n)$ such that
$\omega^n$ has volume one. In particular, as a vector space,
$H^*(\CP^n)$ has the base $1,\omega,\cdots,\omega^n$.
We denote by $t_m^a$ ($a=0,\cdots,n$)
the coupling constant for the observable $\sigma_m(\omega^a)$.
Using the Bott residue formula, we can see
$c_r(\CP^n)={n+1\choose r}\omega^r$.
In particular we have
\beq
\begin{array}{rcl}
\chi(M)\!&=&\!n+1,\\[0.2cm]
\int_{M}c_1(M)\omega^a\omega^b\!&=&\!(n+1)\delta_{b,n-1-a},\\[0.21cm]
\int_{M}c_1(M)c_{n-1}(M)\!&=&\!\frac{(n+1)^2n}{2}.
\end{array}
\eeq
These are enough to see
\beqa
\lefteqn{
L_0=\sum_{m=0}^{\infty}\sum_{a=0}^{n}
\left(\,m+a+\mbox{$\frac{1-n}{2}$}\,\right)
\tilt_m^a\frac{\partial}{\partial t_m^a}
+(n+1)\sum_{m=1}^{\infty}\sum_{a=0}^{n-1}
m\tilt_m^a\frac{\partial}{\partial t_{m-1}^{a+1}}}
\hspace{2cm}\nonumber\\
&&\hspace{0.5cm}+\frac{n+1}{2\lambda^2}\sum_{a=0}^{n-1}t_0^at_0^{n-1-a}
-\frac{1}{48}(n^2-1)(n+3),
\eeqa
where $\lambda$ is the string coupling constant.

\vspace{0.6cm}
{\sc The Grassmannian Model}

\vspace{0.3cm}
The complexgrassmann manifold $M=G(k,N)$
is the space of $k$-dimensional subspaces of $\C^N$.
The cohomology ring of this space is conveniently described
in terms of the Schubert cycles.
For each sequence $a=a_1,\cdots,a_k$ of integers with
$N-k\geq a_1\geq\cdots\geq a_1\geq 0$,
we take a subspace $W_a\subset\C^N$ with the base
$e_{N-k-a_1+1},e_{N-k-a_2+2},\cdots,e_{N-a_k}$ where
$\{e_i\}_{i=1}^N$ is the standard base of $\C^N$.
The orbit through $W_a$ of the group of $N\tms N$
upper triangular regular matrices
is a cell $C_a\subset M$ whose closure $\overline{C_a}$ is a submanifold
of codimension $|a|=a_1+\cdots +a_k$ (the Schubert cycle).
Let $\Omega_a\in H^{2|a|}(M)$ be the Poincar\'e dual of $\overline{C_a}$.
It is known that $\{\Omega_a\}_a$ is a base of the vector space $H^*(M)$.
Hence,
\beq
\chi(M)={N\choose k}.
\eeq
The ring structure is also well known (the Schubert calculus \cite{GH}).
In particular, we have
\beq
\int_M\Omega_a\Omega_b=\delta_{a_1}^{N-k-b_k}\cdots\delta_{a_k}^{N-k-b_1}.
\eeq
In addition, the generator $\omega=\Omega_{1,0,\cdots,0}$ of $H^2(M)$
satisfies
\beq
\omega\wedge\Omega_{a_1,\cdots,a_k}
=\sum_{i=1}^k\Omega_{a_1,\cdots,a_i+1,\cdots, a_k},
\eeq
in which we put $\Omega_{a_1,\cdots,a_i+1,\cdots,a_k}=0$ unless
$a_{i-1}\geq a_i+1\geq a_{i+1}$.
Since $\omega$ is the first chern class
of ``the tautological quotient bundle''
(whose fibre at $W\in M$ is the quotient space $\C^N/W$),
we can easily see
\beq
c_1(M)=N\omega.
\eeq
Using the Bott residue formula, we find that
\beq
\int_Mc_1(M)c_{\dim M-1}(M)=\frac{k}{2}(N^2-N+2-2k){N\choose k}.
\eeq
Denoting by $t_n^a$ the coupling constant for $\sigma_n(\Omega_a)$,
we obtain the expression
\beqa
L_0\!&=&\!\sum_{n=0}^{\infty}\sum_a
\left(\,n+|a|+\mbox{$\frac{1-k(N-k)}{2}$}\,\right)
\tilt_n^a\frac{\partial}{\partial t_n^a}
+N\sum_{n=1}^{\infty}\sum_{i=1}^k\sum_a
n\tilt_n^{a_1,\cdots,a_k}
\frac{\partial}{\partial t_{n-1}^{a_1,\cdots,a_i+1,\cdots,a_k}}\\
&&+\frac{N}{2\lambda^2}\sum_{i=1}^k\sum_a
t_0^{a_1,\cdots,a_i-1,\cdots,a_k}t_0^{N-k-a_1,\cdots,N-k-a_1}
-\frac{1}{48}(kN^2-3k^2+2k-3){N\choose k},\nonumber
\eeqa
where each sum $\sum_a$ is over the sequences
for which the summands are defined.

\newcommand{\NP}{Nucl. Phys.\,}
\newcommand{\PL}{Phys. Lett.\,}
\newcommand{\CMP}{Commun. Math. Phys.\,}

\end{document}